**Astrogenomics: big data, old problems, old solutions?**

Aaron Golden,[1,2,]* S. George Djorgovski,[3,4] John M. Greally[1,2]

[1]Center for Epigenomics, [2]Division of Computational Genetics, Department of Genetics, Albert Einstein College of Medicine, Bronx, NY

[3]Center for Advanced Computing Research, California Institute of Technology, Pasadena, CA

[4]Department of Astronomy, California Institute of Technology, Pasadena, CA

*Corresponding author:

Albert Einstein College of Medicine, 1301 Morris Park Avenue, Bronx, NY 10461 USA

aaron.golden@einstein.yu.edu

+1 718 839 7219  telephone, +1 718 678 1016 fax

The ominous warnings of a 'data deluge' in the life sciences from high-throughput DNA sequencing data are being supplanted by a second deluge, of clichés bemoaning our collective scientific fate unless we address the genomic data 'tsunami'. It is imperative that we explore the many facets of the genome, not just sequence but also transcriptional and epigenetic variability, integrating these observations in order to attain a genuine understanding of how genes function, towards a goal of genomics-based personalized medicine. Determining any individual's genomic properties requires comparison to many others, sifting out the specific from the trends, requiring access to the many in order to yield information relevant to the few. This is the central big data challenge in genomics that still requires some sort of resolution. Is there a practical, feasible way of directly connecting the scientific community to this data universe?

The best answer could be in the stars overhead. About two decades ago, astronomers faced a similar challenge – new digital imaging detectors such as charge-coupled devices (CCDs) and increasingly sophisticated observatories created its own data avalanche, in the era of floppy disks and megabase hard drives. Data acquisition in astronomy has been growing exponentially ever since, with a Moore's Law-like doubling time of 12-18 months [1]. While neither astronomy nor genomics is unique in encountering a big data predicament, astronomy was one of the first scientific disciplines to recognize and tackle these issues effectively, with lessons for today's challenges in genomics.

Two developments facilitated progress by the astronomers. First, the astronomical community recognized the benefits of settling on a common image standard, the Flexible Image Transportation System (FITS), which struck the right balance between universality and simplicity, and has been used as a world-wide astronomical standard ever since [2]. The second important development was the emergence of publicly accessible data archives from all NASA missions that encouraged data re-use and federation. Thus, the astronomical community was primed to use web-based, distributed databases.

As the quality and complexity of data increased, the pressure increased for 'reducing the data', or the processing of raw images to extract specific measurements, encouraging an approach based around a hierarchy of data types and derived data products. The apex of the hierarchy, such as the number of galaxies in a field of view as a function of flux, typically takes up a small fraction of storage space compared to the numerous raw image files from which they are derived. Metadata describing the observational data. containing information about how the images were acquired and processed to generate the reduced image data, were also recognized to be essential, allowing lasting use of the observational data and their integration with other astronomical datasets.

As the data volumes grew far beyond what could be effectively analyzed or downloaded by a scientist on a personal computer, the need emerged for a different paradigm for data access, sharing, and analysis. The advent of the World Wide Web provided a suitable platform for these tasks.

*The Virtual Observatory.*

As a consequence, around the start of the new millennium, representatives of the worldwide astronomical community initiated the development of the Virtual Observatory (VO) framework. This was envisioned as a complete, distributed, web-accessible research environment for astronomy with massive and complex data sets, connecting in a user-transparent manner the data assets, computational resources, tools and even literature [3-6]. The concept was embraced by the astronomical community worldwide, with national and regional VOs unified in the International Virtual Observatory Alliance (IVOA; http://ivoa.net). Similar ideas and frameworks were developed around the same time in other fields, in what is sometimes referred to as the cyberinfrastructure movement [7].

A key idea behind the VO concept is that whereas the individual data repositories remain the responsibility of contributing groups, observatories, and space missions (the people best qualified to curate the data), the VO framework assures their interoperability through a set of common standards, formats, and protocols, enabled by associated metadata. The individual data holdings can then be 'registered' within the VO framework, documented with the proper metadata, and their access and subsequent analysis by various astronomer 'clients' facilitated by the VO's common standards implemented by each participating data node. Thus the entire federated data ecosystem was designed to grow, with interactivity kept both manageable and scalable. It enabled easy data sharing and re-use, whether mandated by the funding agencies, or performed by data producers wanting to see their data used.

The VO approach also enabled scientists less skilled in programming to explore the data. This was possible because the design of the VO emphasized interoperability, creating standards for data and application programming interface (API) management and implementation. The rigid enforcement by the VO of a structured approach to a universal vocabulary of metadata types in observational space was also an essential step. Today, the astronomy community has an infrastructure that fosters the development of tools to implement complex searches across diverse and disparate data archives. Exploration of the skies using the VO can be performed either locally on your desktop computer, or remotely on, for example, NASA supercomputing resources. The

VO's inherent flexibility combined with its strong sense of community ensures its continuing success today as a global 'big data' informatics project.

*Back to Earth.*

So how do we make the jump from stars to genomes? If you replace telescopes with massively parallel sequencers, photometric filters with molecular assays, image processing pipelines with bioinformatics workflows, and multi-wavelength astronomy with integrative genomics, many of the parallels hold up. Perhaps even more pertinently, both endeavors use reference frames – for sequencing, the chromosomal coordinates within the genome, for astronomy, celestial coordinates on the sky. Thus as the apex of molecular observations can be thought of as the expression level of a gene or the methylation of a cytosine at their specific genomic locations, in a similar way the result of an astronomical observation could be the absolute flux measured in visible light of a star at a given location, which in all cases is a tuple with a measurement, a location reference and a specific annotation(s). The challenges inherent to genomics include the relatively greater number and types of sequencing systems compared with astronomical observatories, and the recognized potential for technical and biological variation to influence sequencing-based assays, making it even more critical that there is a universal, standardized framework for data sharing and access in genomics. Can the Virtual Observatory still be a viable model for the life sciences?

*Celestial navigation for the good ship* Genomics.

The essential first step for genomics will involve bringing automated bioinformatic workflows progressively closer to the sequencer itself. Astronomy's equivalent is the dedicated computing hardware and data processing pipelines optimized for a given instrument that make the data analysis-ready in real time. This becomes a critical issue as the data rates continue to explode; for example, the planned radio telescope, Square Kilometer Array (SKA; http://www.skatelescope.org), is anticipated to generate raw data at a rate of ~ 4.2 Petabytes/second, immediately processed and reduced to science-grade data products at a rate of ~0.5-10 Petabytes/day. A data stream from a DNA sequencing platform could likewise be processed in real time, immediately deleting the raw sequence reads, generating as the final output the reduced, aligned or assembled data with a substantially diminished data footprint. A major challenge for the young field of genomics is going to be community acceptance of automated bioinformatic workflow components, as there is often a lack of consensus about the best of choices of tools such as aligners or variant callers.

Assuming this initial processing hurdle can be overcome, the genomic data products could be federated in a manner akin to the VO model, with disparate repositories hosting the reduced, annotated datasets and associated metadata connected into a centralized 'registry', accessed by means of a common suite of tools. While there have been some encouraging steps towards this goal from the modENCODE consortium [8] using the InterMine data warehousing system [9], they do not fully recapitulate the VO model as all data reside in a single repository. Nevertheless, the modENCODE researchers established that life scientists are indeed capable of joining forces to create this VO-like infrastructure. The modENCODE example however falls short of the astronomical community's requirement for global interoperability and imposition of standards for both data and metadata..  Such rigidity may be considered excessive or premature for genomics, but is a critical foundation for the development of tools to implement multidimensional searches across diverse genomic data archives.  The absence of such a common cyberecosystem in the life sciences is stifling the community's potential productivity.

*'Now is the time' – the time for the funding agencies to flex muscle.*

So these are the principal lessons from the astronomers – the genomics community needs to come to agreement on common interoperability standards, allowing data archives and analysis tools to be developed, and to define and adopt metadata standards that provide researchers with the ability both to provenance and to combine individual data sources.  How does one encourage the diverse genomics community to coalesce, and what are the conditions needed to foster an atmosphere in which a cyberecosystem can develop?

Astronomy offers a further lesson – appropriate conditions are fostered by funding agencies.  An influential National Academy of Sciences report [10] prompted US funding agencies (NSF and NASA) to embrace the concept of the VO from the outset.  For genomics in the USA, the National Institutes of Health (NIH) could be reasonably expected to define requirements for responsible use of its funds.  We note that this is not without precedent at the NIH: the heavily-criticized [11] caBIG program of the National Cancer Institute (NCI) had original goals of interoperability of cancer data that were exemplary from the perspective of creating a VO-like framework.

We believe that the NIH needs to be substantially more active in taking advantage of its position of power, coordinating requests for information (RFIs) from members of the genomics community to guide the design of national cyberinfrastructure based around the proven principles of the VO. They could insist on the structured deposition of all genomics data generated from NIH-funded projects to a federally resourced cyberinfrastructure distributed throughout the USA. This creates the VO-like environment within which what may be described as 'evolutionarily convergent' software development can take place, driven by the research community, following NIH-mandated interoperability standards, with the focused goal of developing more complex tools to enhance knowledge discovery.

*And data analysis for all...*

When the idea of the VO was being developed, the developers took inspiration from the then current model of successfully tackling a problem of coordination of standards, the World Wide Web Consortium (W3C). Today, the lessons of the VO should be used to guide the genomics community to meet the challenge of allowing genomics data to be explored to the fullest possible extent. The VO-like outcome of tools generating intuitive visual data representations will expand the analysis of genomics data beyond trained programmers to the biologists and clinicians traditionally unable to perform such explorations independently. With the creation of a cyberecosystem in which evolutionarily convergent software development can take place, the risk of developing dead end software is diminished, the security of highly sensitive sample data is enhanced, and the big data deluge can be stemmed, to the benefit of all.

**REFERENCES**

1. Szalay A, Gray J: **The world-wide telescope.** *Science* 2001, **293:**2037-2040.

2. Wells DC, Greisen EW, Harten RH: **FITS - a Flexible Image Transport System.** *Astronomy and Astrophysics Supplement* 1981, **44:**363.

3. Brunner, R., Djorgovski, S. G., and Szalay, A. (eds.). **Virtual Observatories of the Future.** *Astron. Soc. Pacific Conf. Ser*. 2001, **225**. Provo, UT: Astronomical Society of the Pacific.

4. Djorgovski SG: *Towards the National Virtual Observatory: A report prepared by the National Virtual Observatory Science Definition Team,* 2002 *http://www.us-vo.org/sdt/*

5. Djorgovski SG, Williams R: **Virtual Observatory: From Concept to Implementation.** *Astron. Soc. Pacific Conf. Ser*. 2005, **345,** 517-532. Provo, UT: Astronomical Society of the Pacific.

6. Hanisch R: **The Virtual Observatory in Transition.** *Astron. Soc. Pacific Conf. Ser*. 2006, **351,** 765-770. Provo, UT: Astronomical Society of the Pacific.

7. Atkins DE, Droegemeier KK, Feldman SI, Garcia-Molina H, Klein ML, Messerschmitt DG, Messina P, Ostriker JP, Wright MH: **Revolutionizing Science and Engineering Through Cyberinfrastructure: Report of the National Science Foundation Blue-Ribbon Advisory Panel on Cyberinfrastructure.** 2003 http://www.nsf.gov/cise/sci/reports/atkins.pdf

8. Contrino S, Smith RN, Butano D, Carr A, Hu F, Lyne R, Rutherford K, Kalderimis A, Sullivan J, Carbon S, et al: **modMine: flexible access to modENCODE data.** *Nucleic Acids Res* 2012, **40:**D1082-1088.

9. Smith RN, Aleksic J, Butano D, Carr A, Contrino S, Hu F, Lyne M, Lyne R, Kalderimis A, Rutherford K, et al: **InterMine: a flexible data warehouse system for the integration and analysis of heterogeneous biological data.** *Bioinformatics* 2012, **28:**3163-3165.

10. *Astronomy and Astrophysics in the New Millennium.* The National Academies Press; 2001.

11. Califano A, Chinnaiyan A, Duyk GM, Gambhir SS, Hubbard T, Lipman DJ, Stein L, Wang JY: **An assessment of the impact of the NCI Cancer Biomedical Informatics Grid (caBIG).** http://deainfo.nci.nih.gov/advisory/bsa/bsa0311/caBIGfinalReport.pdf; 2011.

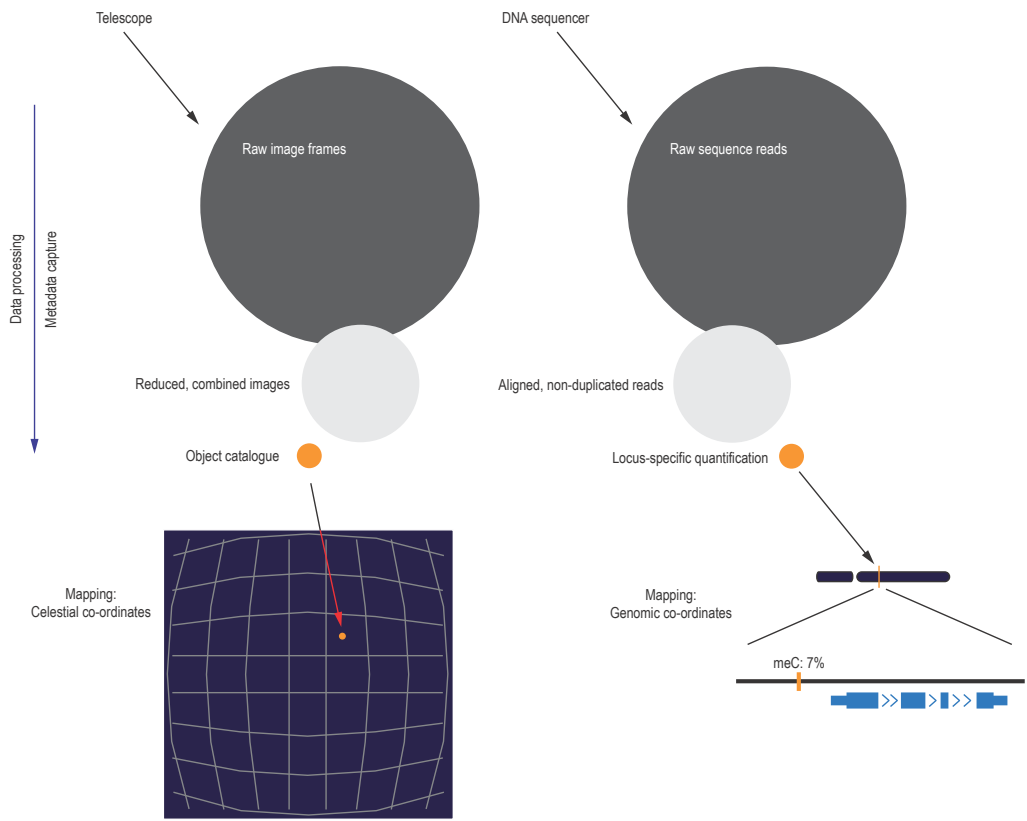

**FIGURE LEGEND**

The parallels between astronomy and DNA sequencing can be illustrated in terms of the extremely large initial data sets generated that are reduced in size and increased in information content by analytical algorithms. The processed information in each case can be provided with co-ordinates within a reference frame, allowing registration and integration of different types of data.